In-plane spin–orbit torque magnetization switching and its detection using the spin rectification effect at sub-GHz frequencies

Motomi Aoki[1], Ei Shigematsu[1], Masayuki Matsushima[1], Ryo Ohshima[1], Syuta Honda[2], Teruya Shinjo[1], Masashi Shiraishi[1] and Yuichiro Ando[1,†]

[1]Department of Electronic Science and Engineering, Kyoto University, Kyoto, Kyoto 615-8510, Japan

[2]Department of Pure and Applied Physics, Kansai University, Suita, Osaka 564-8680, Japan

**Corresponding authors**

[†]**Yuichiro Ando** Address: A1-226, Kyodai Katsura, Nishigyo-ku, Kyoto, Kyoto, Japan

    Tel.: +81-75-383-2356; Fax: +81-75-383-2275

    E-mail: ando@kuee.kyoto-u.ac.jp

**Abstract**

In-plane magnetization reversal of a permalloy/platinum bilayer was detected using the spin rectification effect. Using a sub-GHz microwave frequency to excite spin-torque ferromagnetic resonance (ST-FMR) in the bilayer induces two discrete DC voltages around an external static magnetic field of 0 mT. These discrete voltages depend on the magnetization directions of the permalloy and enable detection of the in-plane magnetization reversal. The threshold current density for the magnetization reversal is around 10–20 MA/cm$^2$, the same order as for known spin–orbit torque (SOT) switching with in-plane magnetization materials. The magnitude of the signal is the same or larger than that of the typical ST-FMR signal; that is, detection of magnetization switching is highly sensitive in spite of deviation from the optimal ST-FMR condition. The proposed method is applicable to a simple device structure even for a small ferromagnetic electrode with a width of 100 nm.

## I. INTRODUCTION

Magnetization switching using spin–orbit torque (SOT) [1–7] has attracted much research interest because it enables fast, low-power, and high-endurance write operations in nonvolatile magnetic memories such as magnetoresistive random access memory (MRAM). The large spin Hall angle $\theta_{SHE}$ of spin–orbit materials makes them essential for highly efficient magnetization switching via SOT. Therefore, quantitative investigations of $\theta_{SHE}$ have been conducted on a wide variety of materials including metals [8–14], semiconductors [15–18], topological insulators [19–28], and even light-element materials [29–32]. Spin torque ferromagnetic resonance (ST-FMR) is one of the main methods for this, where $\theta_{SHE}$ is evaluated from the line shape of the DC voltage signals generated by the spin rectification effect (SRE) [33–35]. Typical ST-FMR signals consist of two components, symmetric and antisymmetric Lorentzian functions. In the simplest case, $\theta_{SHE}$ is calculated from the ratio of their magnitudes. However, the Oersted field induced by an applied microwave generates both symmetric and antisymmetric components whose magnitudes are sensitive to the phase difference between the Oersted field and the magnetic moment $M$ [34,36–38]. Furthermore, one should consider the contributions of the anomalous Nernst effect (ANE), the spin Seebeck effect (SSE), and the inverse spin Hall effect (ISHE) induced by spin pumping [39–41]. The field-like torque also generates an antisymmetric component that becomes pronounced when the interfacial Rashba effect is significant. Various approaches have been proposed to distinguish SOT-related FMR signals from unwanted effects. These methods include phase-sensitive detection, comparing different ferromagnetic metals, using signal dependence on the thickness of the ferromagnetic layer, and combining with optical detection [42,43]. However, their procedures for device fabrication, measurement, and analysis are complicated.

Most ST-FMR studies aim to achieve ultrafast and low-power magnetization switching. Therefore, material search using SOT magnetization switching is a determinative method. The device structure for magnetization switching, however, is more complicated than that for ST-FMR, which impedes the search for a wide variety of spin–orbit materials. Magnetization switching in perpendicular magnetization materials can be detected via the anomalous Hall effect (AHE). Confirming magnetization switching in this case needs fabrication of two nonmagnetic electrodes for measuring the Hall voltage of the ferromagnetic electrode whose polarization corresponds to the magnetization direction. Applying the AHE to in-plane magnetization materials is generally difficult. Therefore, additional spin valves such as magnetic tunnel junctions (MTJs) and giant magnetoresistive (GMR) devices are commonly fabricated. [2,44,45]. It is much preferred to detect magnetization switching of a single ferromagnetic layer with a simple device structure.

In this study, we detect in-plane magnetization switching of a single ferromagnetic layer on spin–orbit materials by applying sub-GHz microwaves. The new method, hereafter called "low-frequency ST-FMR" (LFST-FMR), has the same device structure as that of ST-FMR and does not require fabricating additional electrodes or spin valves. By only applying sub-GHz microwaves to an electrode without a magnetic field, we determine the magnetization direction as the polarity of a DC voltage. Surprisingly, the detection sensitivity is higher than that of ST-FMR. Therefore, magnetization switching can be detected even for a ferromagnetic electrode as small as 100 nm in width, less than one-tenth of the minimum spatial resolution of optical methods using the magneto Kerr effect [46].

## II. SAMPLE FABRICATION AND EXPERIMENTAL PROCEDURE

Figures 1(a) and 1(b) show schematics of the device structure and the equivalent electric circuit in this study. First, rectangular-shaped MgO (2 nm) / $Ni_{80}Fe_{20}$ (Py, 4 nm) / Pt (15 nm) layers were deposited on the MgO(100) substrate using electron beam lithography and electron beam deposition. The sample was exposed to the air, and then a 7-nm-thick $SiO_2$ layer was deposited on the MgO layer via RF magnetron sputtering. After liftoff of the $SiO_2$/MgO/Py/Pt layers, a rectangular $SiO_2$/MgO/Py ferromagnetic electrode was fabricated using electron beam lithography and argon ion ($Ar^+$) milling. The $Ar^+$ milling was stopped after milling of the 4-nm-thick Pt layer to keep an 11-nm-thick Pt channel. Finally, a Au (100 nm) / Ti (3 nm) coplanar wave guide was fabricated using electron beam lithography and electron beam deposition. The length of the Py layer along the $y$ direction, $L_y$, was varied from 1 to 20 μm and that along the $x$ direction, $L_x$, was varied from 0.1 to 20 μm. In the ST-FMR studies, a static magnetic field was applied in the film plane with a varying field orientation angle $\theta_H$ as shown in Fig. 1(a). (See the Supplemental Material for the detailed fabrication procedures.) Microwave radiation with frequency $f_{MW}$ was applied using an analog commercial signal generator (KEYSIGHT N5173B EXG) to excite the FMR of the Py layer, and a DC voltage $V_{DC}$ was measured using a nanovoltmeter (Keithley Nanovoltmeter

2182A). In the magnetization switching study, a function generator (Agilent 33622A) was used to apply a pulse-shaped charge current (pulse current). All the measurements were carried out at room temperature.

Figure 1(c) shows typical ST-FMR signals, i.e., $V_{DC}$ as a function of the external magnetic flux density $B_{ext}$, at $\theta_H = 45°$ for various values of $f_{MW}$. The values of $L_x$ and $L_y$ are 500 nm and 20 μm, respectively, one or two orders lower than those of typical ST-FMR measurement devices. The microwave power $P_{MW}$ was 10 dBm. Clear ST-FMR signals were obtained both at positive and negative $B_{ext}$ for up and down sweeps. Whereas similar ST-FMR signals were obtained for $f_{MW} = 6$–9 GHz, the resonance field $B_{res}$ gradually approaches 0 mT with decreasing $f_{MW}$. The ST-FMR signal $V_{DC}(\theta_M, B_{ext})$ is expressed as

$$V_{DC}(\theta_M, B_{ext}) = \frac{1}{4}\frac{dR}{d\theta_M}\frac{\gamma I_{MW}\sin\theta_M}{\Gamma 2\pi(df_{MW}/dB_{ext})|_{B_{ext}=B_{res}}}\left(S\frac{\Gamma^2}{\Gamma^2+(B_{ext}-B_{res})^2} + A\frac{\Gamma(B_{ext}-B_{res})}{\Gamma^2+(B_{ext}-B_{res})^2}\right), \quad (1)$$

where $R$ is the resistance of the Py/Pt device, $I_{MW}$ is the rf current through the microstrip, $\theta_M$ is the angle of the magnetization rotation axis from the +$y$ direction as shown in Fig. 1(b), $S$ and $A$ are the coefficients of the symmetric and antisymmetric Lorentzian, respectively, and $\Gamma$ is the half-width at half maximum of the Lorentzian function. We obtained $B_{res}$ and $\Gamma$ by fitting Eq. (1). For the in-plane ferromagnetic films, the FMR condition is the Kittel formula [12]:

$$f_{MW} = \frac{\gamma}{2\pi}\sqrt{(B_{res} + (N_x - N_y)\mu_0 M_s)(B_{res} + (N_z - N_y)\mu_0 M_s)}, \quad (2)$$

where $M_s$ is the saturation magnetization; $\gamma$ is the gyrometric ratio; and $N_x, N_y$, and $N_z$ are the demagnetization factors along the $x$, $y$, and $z$ axes, respectively. The fitting using Eq. (2) is shown in Fig. 1(d), which yields $(N_z - N_y)\mu_0 M_s = 0.58$ T and $(N_x - N_y)\mu_0 M_s = 4.2\times10^{-16}$ T. Figure 1(e) shows $\Gamma$ for the FMR signals as a function of $f_{MW}$, confirming a linear relationship. The intercept of the $\Gamma$ axis, caused by non-uniformity of the ferromagnetic layer, is 0.2 mT, indicating good quality of the Py/Pt bilayer structure.

**III. SUB GHz LFST-FMR SIGNALS**

Now we focus on the ST-FMR signals at low $f_{MW}$, namely, LFST-FMR signals. Figure 2(a) shows the ST-FMR signals with various $f_{MW}$ values at $\theta_H = 45°$. For $f_{MW} = 5$ GHz, clear ST-FMR signals similar to those of Fig. 1(c) were obtained around $B_{res} = \pm45$ mT for both up and down sweeps. $B_{res}$ was decreased by decreasing $f_{MW}$ and finally reached around 0 mT for $f_{MW} \leq 2$ GHz, according to the Kittel formula as shown in Fig. 1(e).

The signal shapes for up and down sweeps are almost identical for $f_{MW} \geq 2$ GHz. In contrast, the field range of the two ST-FMR signals corresponding to positive and negative $B_{res}$ are superimposed on each other for $f_{MW} \leq 1$ GHz. Figure 2 shows the enlarged ST-FMR signals around 0 mT at $f_{MW}$ = (b) 10 MHz, (c) 200 MHz, (d) 1 GHz, (e) 2 GHz, (f) 3 GHz, and (g) 5 GHz. For $f_{MW}$ = 5 GHz, small hysteresis with a magnitude of 20 nV was obtained around 0 mT. This might be due to the nonresonant SRE or ANE [47,48]. As $f_{MW}$ decreased, the magnitude of the hysteresis increased, and the steep changes in $V_{DC}$ around ± 2 mT due to the magnetization switching became pronounced. Here we note that the upper and lower $V_{DC}$ values in the hysteresis correspond to the magnetization directions along the +y and –y directions, respectively. This means that we can detect the magnetization direction by measuring $V_{DC}$ with low-$f_{MW}$ microwaves. The hysteresis was detected even for $f_{MW}$ < 10 MHz and came under the detection limit for $f_{MW}$ < 1 MHz. It should be noted that magnetization is almost aligned along the y direction around 0 mT because of the shape anisotropy, indicating that the LFST-FMR signals, i.e., the rectangular hysteresis in the $V_{DC}$–$B_{ext}$ curves, were generated with the magnetization along y. To confirm this, ST-FMR signals were measured at $\theta_H = 0°$. The results are displayed in Fig. 3. The usual ST-FMR signals at high $B_{ext}$ disappeared for $f_{MW} \geq 2$ GHz, which is expected from Eq. (1). In contrast, considerable rectangular LFST-FMR signals were still detected for $f_{MW} < 2$ GHz. The rectangular shapes were slightly modulated from those of $\theta_H=45°$ because of $\theta_H$-dependent magnetization switching. Figure 4 shows the $f_{MW}$ dependence of the hysteresis magnitude $\Delta V_0$, which is the difference between the upper and lower $V_{DC}$ values of the LFST-FMR signals at $B_{ext}$ = 0 mT for $\theta_H = 0°$ and 45°. $\Delta V_0$ shows a maximum around $f_{MW}$ = 50–500 MHz both for $\theta_H = 0°$ and 45°. Surprisingly, $\Delta V_0$ for 50–500 MHz is comparable to or greater than the magnitude of the ST-FMR signals at $\theta_H =45°$ for $f_{MW} > 3$ GHz. The possible origin of such a large $\Delta V_0$ will be discussed in Section V.

**IV. DETECTION OF MAGNETIZATION SWITCHING**

We now shift our focus to the main result of this work. Figure 5(a) shows the experimental procedure for demonstrating SOT magnetization switching. First, (1) the magnetization direction was set to the initial direction (+y or –y) by applying a sufficient magnetic flux density $B_{SET}$ with $\theta_H = 0°$. Then, (2) $B_{SET}$ was set to 0 mT, and LFST-FMR was used to measure the DC voltage in the initial magnetization direction, $V_{DC1}$. Then, (3) the circuit switch was changed from A to B to protect the signal generator from the pulse voltage generated by the pulse generator (see Fig. 1(b)). After the circuit was changed, (4) a pulse voltage was applied under a small magnetic flux density $B_{PLS}$. Finally, (5) the switch was changed from B to A followed by (6) measuring $V_{DC2}$ at 0 mT to compare with $V_{DC1}$. The procedure was repeated with a different pulse voltage. Steps (1) and (2) are optional and can be skipped except at the beginning of the measurements. Figure 5(b) shows a hysteresis SOT-switching signal, $\Delta V = (V_{DC2} - V_{DC1})$, as a function of pulse current density $J_{PLS}$ in the Pt layer. $L_x$, $L_y$, $f_{MW}$, and $P_{MW}$ were 500 nm, 20 μm, 200 MHz, and 10 dBm, respectively. $B_{SET}$ was +300 (–300) mT for the up (down) $J_{PLS}$ sweep, indicating that the initial magnetization direction was $+y$ ($-y$) for the up (down) sweep. Steps (1)–(6) were all repeated for each measurement. In Fig. 5, $B_{PLS}$ was precisely adjusted to 0 mT by monitoring the Hall device. The pulse width of $W_{PLS}$ was 1 ms. Because $\Delta V$ is defined as the difference $V_{DC2} - V_{DC1}$, the raw $\Delta V - J_{PLS}$ curve does not show the typical hysteresis of SOT switching, as shown in the inset of Fig. 5(b). Therefore, as shown in the main panel of Fig. 5(b), $V_{DC2}$ in the magnetization direction along $+y$ was set to 0 μV for easy analysis. A clear hysteresis with steep change $\Delta V$ around $J_{PLS} = \pm 15$ MA/cm$^2$ was observed, a typical feature of SOT magnetization switching. The magnitude of the hysteresis was 600 nV, comparable to $\Delta V_0$ in Fig. 4. Next, we skipped the optional steps (1)–(2), i.e., negative (positive) $B_{SET}$ was applied once before starting the up (down) sweep measurements. A similar hysteresis was obtained as shown in Fig. 5(c).

Hereafter, we show the $\Delta V - J_{PLS}$ curves only for the up sweep because those of the down sweep are the same in the following discussions. Figure 5(d) shows the $\Delta V - J_{PLS}$ curve for various $f_{MW}$ values at $\theta_H = 0°$. Whereas the $\Delta V - J_{PLS}$ curve shows almost the same behavior for 10 MHz ≤ $f_{MW}$ ≤ 200 MHz, its magnitude gradually decreases with increasing $f_{MW}$ and is almost zero above 2 GHz, in agreement with the LFST-FMR signals in Fig. 3(a) and Fig. 4. $\Delta V - J_{PLS}$ curves for various pulse widths $W_{PLS}$ are shown in Fig. 5(e). The threshold value of $J_{PLS}$ was 10 MA/cm$^2$ for $W_{PLS} \geq 500$ μs, and it increased with decreasing $W_{PLS}$. Several step-like shapes also appear during the magnetization switching. Because $L_y$ was 20 μm, which was considerably long compared with $L_x$, domain wall propagation along the $y$ direction is expected to be dominant in the magnetization switching process. In this situation, the shape of the $\Delta V - J_{PLS}$ curves strongly depends on the velocity of the domain wall propagation. Assuming a typical domain wall velocity of the Py layer ($v < 100$ m/s), at least 200 ns is needed to propagate the domain wall through the 20-μm-long Py layer [49–51]. The heat effect also contributes to modulating the

ΔV–$J_{PLS}$ curves [52,53]. To evaluate the SOT switching time accurately, $L_y$ should be decreased to a practical device size. Figures 5(f) and 5(g) show the $L_x$ dependence of the ΔV–$J_{PLS}$ curves and the normalized ones. The magnitude of the ΔV–$J_{PLS}$ curves monotonically increases with increasing $L_x$ as shown in the inset of Fig. 5(f), indicating that a large $L_x$ is needed to obtain a large signal. The switching feature was made steeper by decreasing $L_x$ because the domain structure became more complicated for large $L_x$, resulting in complicated domain wall motion.

## V. Micromagnetic simulation of ST-FMR and field-induced FMR

In this section, we discuss the possible origin of the obtained DC voltage in LFST-FMR. There are several differences in conditions between LFST-FMR and typical ST-FMR measurements. First, $B_{ext}$ was 0 mT during the FMR measurements in LFST-FMR. Therefore, the magnetization is expected to align along the *y* direction owing to the shape anisotropy. Second, the microwave frequency was sub-GHz, one or two orders smaller than typical frequencies. Third, the AC magnetic field was along the *y* direction, parallel to the magnetization direction. $V_{DC}$ due to the spin rectification effect is

$$V_{DC} = \langle I_{MW}\sin(2\pi f_{MW}t)\ B_{MW}\frac{dR}{dB_{ext}}\sin(2\pi f_{MW}t - \varphi)\rangle,$$

where $\langle\ \rangle$ denotes the time average, $\varphi$ is the phase difference between *M* and alternating current in the Py/Pt device, *t* is the time, and $B_{MW}$ is the magnetic flux density generated by the microwave. $\frac{dR}{dB_{ext}}$ is generally expected to be negligible at $\theta_H = 0°$ because $B_{MW}$ is parallel to the magnetization direction and the precession axis is along the *y* direction. However, the magnitude of the LFST-FMR in Fig. 3 is unexpectedly comparable to or higher than that of the typical ST-FMR in Figs. 1 and 2. To find the origin of such a large signal, we first measured the magnetoresistance (MR) of the device by applying $B_{ext}$ at $\theta_H = 0°$ and a direct current of 1 mA. Figure 6(a) shows the MR ratio as a function of $B_{ext}$. A direct current was applied along the +*x* direction. Triangular hysteresis signals were obtained with a switching field of $B_{ext} = 3$ mT, a typical shape for anisotropic magnetoresistance (AMR) with an external magnetic field along the magnetic easy axis. We also measured the MR ratio as a function of $\theta_H$ for the Py/Pt bilayer without device fabrication to investigate the AMR magnitude in the Py/Pt bilayer structure. The applied magnetic field was 100 mT, sufficiently high to align the magnetization along $B_{\text{ext}}$. Whereas a clear sine shape was obtained as shown in Fig. 6(b), the MR ratio was only 0.07 %, approximately one order of magnitude smaller than that of typical Py, owing to the considerable parallel conduction of the Pt layer. In practical devices, the MR ratio is much smaller than that in Fig. 6(b) because the parasitic resistance of the Pt leads connected in series to the Py/Pt bilayer electrodes is dominant. Here we note that a finite $\frac{dR}{dB_{\text{ext}}}$ was obtained at 0 mT as shown in Fig. 6(a), which might be due to the tilt of the magnetization especially at edge areas (see the Supplemental Material). $\frac{dR}{dB_{\text{ext}}}$ around $0 \pm 0.5$ mT was $+27 \pm 2$ and $-53 \pm 2$ m$\Omega$/T for the up and down sweep, respectively. The nonzero $\frac{dR}{dB_{\text{ext}}}$ around 0 mT indicates that the resistance of the sample actually changes even for application of low $B_{\text{MW}}$ along the $y$ axis. $B_{\text{MW}}$ is estimated to be 0.43 mT at 10 dBm, which yields a resistance change of $3.7 \times 10^{-5}$ $\Omega$.

From the micromagnetic simulation, the magnetic resonance frequency is calculated to be 1.425 GHz at 0 mT (see Supplemental Material). For $f_{\text{MW}} < 1$ GHz, $\varphi$ due to the SOT and Oersted field is expected to be 90º and 0º, respectively. In this case, $\Delta V_0$ at $B_{\text{ext}} = 0$ mT and $\theta_H = 0º$ due to the Oersted field is simply expressed as

$$\Delta V_0 = I_{\text{MW}} B_{\text{MW}} \frac{dR}{dB_{\text{ext}}} .$$

Because we measured $\frac{dR}{dB_{\text{ext}}}$ for the whole Py/Pt bilayer, we used an alternating current in the whole Py/Pt layer, $I_{\text{MW}}$. Using $I_{\text{MW}} = 1.8 \times 10^{-2}$ A, $B_{\text{MW}} = 0.43$ mT, and $\frac{dR}{dB_{\text{ext}}} = 40$ m$\Omega$/T, $\Delta V_0$ is calculated to be 310 nV, approximately half of the experimental value in Fig. 4. The difference between experimental and theoretical $\Delta V_0$ might be due to the contribution of ferromagnetic resonance. Whereas the ferromagnetic resonance frequency at 0 mT was around 1.425 GHz (see Supplemental Material), it has a finite dispersion according to location in the Py layer. Therefore, several parts of the Py electrode match the ferromagnetic condition, resulting in an enhancement of $\frac{dR}{dB_{\text{ext}}}$ [54]. In this case, $\varphi$ changes from 0º, and an SOT contribution to $\Delta V_0$ is expected.

$V_{\text{DC}}$ can also be generated by the anomalous Nernst effect (ANE) and by the combination of spin pumping and the ISHE. For the ANE, the temperature gradient along the $z$ direction due to microwave absorption thermally activates carriers that flow along the $z$ direction and generate an electric field along the $x$ direction. The ANE DC voltage $V_{\text{ANE}}$ is

$$V_{\text{ANE}} = \alpha_{\text{N}} \frac{L_x \Delta T_{\text{Py}}}{t_{\text{Py}}} ,$$

where $\alpha_N$=4.8 nV/K is the ANE coefficient of Py, and $\Delta T_{Py}$ is the temperature difference between the top and bottom surfaces of the Py layer [55]. $\Delta T_{Py}$ is generally less than several tens of mK at maximum because the Py layer is very thin. Assuming $\Delta T_{Py} = 10$ mK, $V_{ANE}$ is expected to be at maximum 6 nV in this study, which is negligible. For the combination of spin pumping and the ISHE, magnetic precession of the Py layer generates a pure spin current $J_s$ along the $z$ direction, and $J_s$ is converted into a charge current via the ISHE. Assuming the typical mixing conductance in Py/Pt interfaces, $g_r^{\uparrow\downarrow} = 2.31 \times 10^{19}$ m$^{-2}$, the spin current generated by spin pumping is estimated to be $J_s = 1.7 \times 10^{-14}$ J/m$^2$ [56,57]. Using the typical spin Hall angle, $\theta_{SHE} = 0.04$, and spin diffusion length of 10 nm, the DC voltage generated by the ISHE is estimated to be $2.4 \times 10^{-13}$ V, also negligible compared with $\Delta V_0$ in Fig. 4 (see Supplemental Material).

## VI. APPLICATION TO A PRACTICAL DEVICE WITH A SMALL FERROMAGNET

The typical device of a practical MTJ has a diameter of less than 100 nm, which is one order of magnitude smaller than that of our device. Furthermore, Py is not a suitable ferromagnetic material for a high MR ratio in an MTJ. Therefore, we now discuss the possibility of applying LFST-FMR to the small ferromagnetic electrode of a practical device. We first fabricated a sample with $L_x$=100 nm and $L_y$=1000 nm. The LFST-FMR signal is shown in Fig. 7(a). Because $V_{DC}$ depends not on $L_y$ but on $L_x$, a decrease in $L_y$ is possible without significant reduction in $V_{DC}$. We determined $L_y$ for impedance matching and sufficient shape anisotropy. $P_{MW}$ and $W_{PLS}$ were 5 dBm and 1 μs, respectively, because of the smaller device size. A clear rectangular signal was still detected in the $V_{DC}$–$B_{ext}$ curve as shown in Fig. 7(a). $\Delta V_0$ at 0 mT is around 0.15 μV, approximately one-fourth that of the $L_x$=500 nm sample in Fig. 3. Figure 7(b) shows $\Delta V$– $J_{PLS}$ curves for the $L_x$=100 nm device. A varying static magnetic field $B_{PLS}$ was applied during the pulse voltage (see step (3) in Fig. 5(a)) to investigate contributions of the SOT and Oersted field to magnetization switching. Positive (negative) $B_{PLS}$ corresponds to a magnetic field that impedes (assists) the magnetization switching. For $B_{PLS} = 0$ mT, a steep change in $\Delta V$ is obtained at 25 MA/cm$^2$. The switching feature was steeper than in Fig. 5 because $L_y$, which determines the length of the domain wall propagation, was decreased by a factor of 20. Whereas the Oersted field from $J_{PLS}$ is estimated to be less than 1–2 mT, successful switching was obtained even for $B_{PLS} = 2$mT, indicating a non-negligible SOT contribution.

We also propose an enhancement of signal amplitude and a statistical investigation of the magnetization switching characteristics by fabricating several Py electrodes in series on a Pt layer. An increased number of ferromagnetic electrodes is expected to contribute an increase in $V_{DC}$, which enables downsizing of the ferromagnetic electrodes. Figure 7(c) displays an ST-FMR signal for a device equipped with three ferromagnetic electrodes with $L_x = 500$ nm. Clear hysteresis was obtained with $\Delta V_0 = 500$ nV ($P_{MW} = 10$ dBm). Because the length of the Pt lead was increased by increasing the number of Py electrodes in this device, $\Delta V_0$ does not show a linear relationship with the number of Py electrodes. Optimization of the Pt lead structure is desired.

To confirm the applicability of LFST-FMR to other ferromagnetic materials, we fabricated a device with a Co electrode on Pt films. The AMR magnitude of Co is much smaller than that in a Py layer [58,59]. We obtained clear rectangular signals even for the Co electrode as shown in Fig. 7(c). The magnitude of $\Delta V_0$ depends on the charge current density in the ferromagnetic layer and magnitude of $\frac{dR}{dB_{ext}}$. Whereas low $\frac{dR}{dB_{ext}}$ is expected owing to the low AMR, the charge current density in the Co layer is expected to increase owing to the low resistivity of the Co layer compared with that of the Py electrode, resulting in considerable $\Delta V_0$. Because we used a relatively thick Pt layer (15 nm), further enhancement of $\Delta V_0$ is possible by decreasing the thickness of the Pt layer. The coercive force of ferromagnetic electrodes in a practical device is expected to be greater than that of a Py electrode, resulting in $\frac{dR}{dB_{ext}} \approx 0$ [Ω/T] around 0 mT and negligible $\Delta V_0$. Even in this case, LFST-FMR can be applied using a finite magnetic field during FMR measurements as long as the $R$–$B_{ext}$ curve shows hysteresis behavior. When magnetization is applied, the magnetization starts to tilt from the $y$ axis only when the magnetic field is antiparallel to the magnetization direction. In this situation, a different $\frac{dR}{dB_{ext}}$ value is obtained for $M // +y$ and $M // -y$.

## VII. Conclusion

We have demonstrated in-plane magnetization switching of a permalloy/platinum bilayer induced by SOT using the spin rectification effect with sub-GHz microwaves. Because the magnitude of the LFST-FMR signal is the same or larger than that of a typical ST-FMR signal, a highly sensitive detection of magnetization switching has been realized in spite of deviation from the optimal ST-FMR condition. The proposed method is applicable to a simple device structure even for a ferromagnetic electrode as small as 100 nm wide.

**References**

[1] I. M. Miron, K. Garello, G. Gaudin, P. J. Zermatten, M. V. Costache, S. Auffret, S. Bandiera, B. Rodmacq, A. Schuhl, and P. Gambardella, Nature **476**, 189 (2011).

[2] L. Liu, C.-F. Pai, Y. Li, H. W. Tseng, D. C. Ralph, and R. A. Buhrman, Science **336**, 555 (2012).

[3] L. Liu, O. J. Lee, T. J. Gudmundsen, D. C. Ralph, and R. A. Buhrman, Physical Review Letters **109**, 096602 (2012).

[4] K. Garello, I. M. Miron, C. O. Avci, F. Freimuth, Y. Mokrousov, S. Blügel, S. Auffret, O. Boulle, G. Gaudin, and P. Gambardella, Nature Nanotechnology **8**, 587 (2013).

[5] G. Yu, P. Upadhyaya, Y. Fan, J. G. Alzate, W. Jiang, K. L. Wong, S. Takei, S. A. Bender, L. Te Chang, Y. Jiang, M. Lang, J. Tang, Y. Wang, Y. Tserkovnyak, P. K. Amiri, and K. L. Wang, Nature Nanotechnology **9**, 548 (2014).

[6] K. Garello, C. O. Avci, I. M. Miron, M. Baumgartner, A. Ghosh, S. Auffret, O. Boulle, G. Gaudin, and P. Gambardella, Applied Physics Letters **105**, (2014).

[7] S. Fukami, T. Anekawa, C. Zhang, and H. Ohno, Nature Nanotechnology **11**, 621 (2016).

[8] S. O. Valenzuela and M. Tinkham, Nature **442**, 176 (2006).

[9] T. Kimura, Y. Otani, T. Sato, S. Takahashi, and S. Maekawa, Physical Review Letters **98**, 156601 (2007).

[10] T. Seki, Y. Hasegawa, S. Mitani, S. Takahashi, H. Imamura, S. Maekawa, J. Nitta, and K. Takanashi, Nature Materials **7**, 125 (2008).

[11] Y. Niimi, Y. Kawanishi, D. H. Wei, C. Deranlot, H. X. Yang, M. Chshiev, T. Valet, A. Fert, and Y. Otani, Physical Review Letters **109**, 156602 (2012).

[12] H. L. Wang, C. H. Du, Y. Pu, R. Adur, P. C. Hammel, and F. Y. Yang, Physical Review Letters **112**, 1 (2014).

[13] H. An, Y. Kageyama, Y. Kanno, N. Enishi, and K. Ando, Nature Communications **7**, 1 (2016).

[14] Y.-C. Lau, H. Lee, G. Qu, K. Nakamura, and M. Hayashi, Physical Review B **99**, 064410 (2019).

[15] K. Ando, S. Takahashi, J. Ieda, H. Kurebayashi, T. Trypiniotis, C. H. W. Barnes, S. Maekawa, and E. Saitoh, Nature Materials **10**, 655 (2011).

[16]   K. Ando and E. Saitoh, Nature Communications **3**, 629 (2012).

[17]   M. Koike, E. Shikoh, Y. Ando, T. Shinjo, S. Yamada, K. Hamaya, and M. Shiraishi, Applied Physics Express **6**, 9 (2013).

[18]   A. Yamamoto, Y. Ando, T. Shinjo, T. Uemura, and M. Shiraishi, Physical Review B - Condensed Matter and Materials Physics **91**, 1 (2015).

[19]   A. R. Mellnik, J. S. Lee, A. Richardella, J. L. Grab, P. J. Mintun, M. H. Fischer, A. Vaezi, A. Manchon, E. A. Kim, N. Samarth, and D. C. Ralph, Nature **511**, 449 (2014).

[20]   Y. Shiomi, K. Nomura, Y. Kajiwara, K. Eto, M. Novak, K. Segawa, Y. Ando, and E. Saitoh, Physical Review Letters **113**, 196601 (2014).

[21]   Y. Ando, T. Hamasaki, T. Kurokawa, K. Ichiba, F. Yang, M. Novak, S. Sasaki, K. Segawa, Y. Ando, and M. Shiraishi, Nano Letters **14**, 6226 (2014).

[22]   Y. Fan, P. Upadhyaya, X. Kou, M. Lang, S. Takei, Z. Wang, J. Tang, L. He, L. Te Chang, M. Montazeri, G. Yu, W. Jiang, T. Nie, R. N. Schwartz, Y. Tserkovnyak, and K. L. Wang, Nature Materials **13**, 699 (2014).

[23]   Y. Fan, X. Kou, P. Upadhyaya, Q. Shao, L. Pan, M. Lang, X. Che, J. Tang, M. Montazeri, K. Murata, L. Te Chang, M. Akyol, G. Yu, T. Nie, K. L. Wong, J. Liu, Y. Wang, Y. Tserkovnyak, and K. L. Wang, Nature Nanotechnology **11**, 352 (2016).

[24]   K. Kondou, R. Yoshimi, A. Tsukazaki, Y. Fukuma, J. Matsuno, K. S. Takahashi, M. Kawasaki, Y. Tokura, and Y. Otani, Nature Physics **12**, 1027 (2016).

[25]   Y. Ando and M. Shiraishi, Journal of the Physical Society of Japan **86**, (2017).

[26]   N. H. D. Khang, Y. Ueda, and P. N. Hai, Nature Materials **17**, 808 (2018).

[27]   Y. Liu, J. Besbas, Y. Wang, P. He, M. Chen, D. Zhu, Y. Wu, J. M. Lee, L. Wang, J. Moon, N. Koirala, S. Oh, and H. Yang, Nature Communications **9**, 2492 (2018).

[28]   Z. Chi, Y.-C. Lau, X. Xu, T. Ohkubo, K. Hono, and M. Hayashi, Science Advances **6**, eaay2324 (2020).

[29]   K. Ando, S. Watanabe, S. Mooser, E. Saitoh, and H. Sirringhaus, Nature Materials **12**, 622 (2013).

[30]   R. Ohshima, A. Sakai, Y. Ando, T. Shinjo, K. Kawahara, H. Ago, and M. Shiraishi, (n.d.).

[31] J. B. S. Mendes, O. Alves Santos, L. M. Meireles, R. G. Lacerda, L. H. Vilela-Leão, F. L. A. Machado, R. L. Rodríguez-Suárez, A. Azevedo, and S. M. Rezende, Physical Review Letters **115**, 226601 (2015).

[32] S. Dushenko, H. Ago, K. Kawahara, T. Tsuda, S. Kuwabata, T. Takenobu, T. Shinjo, Y. Ando, and M. Shiraishi, Physical Review Letters **116**, 166102 (2016).

[33] L. Liu, T. Moriyama, D. C. Ralph, and R. A. Buhrman, Physical Review Letters **106**, 1 (2011).

[34] M. Harder, Z. X. Cao, Y. S. Gui, X. L. Fan, and C. Hu, **054423**, 1 (2011).

[35] M. Harder, Y. Gui, and C.-M. Hu, Physics Reports **661**, 1 (2016).

[36] A. Wirthmann, X. Fan, Y. S. Gui, K. Martens, G. Williams, J. Dietrich, G. E. Bridges, and C. M. Hu, Physical Review Letters **105**, 1 (2010).

[37] Y. Zhang, Q. Liu, B. F. Miao, H. F. Ding, and X. R. Wang, Physical Review B **99**, 064424 (2019).

[38] U. Chaudhuri, R. Mahendiran, and A. O. Adeyeye, Applied Physics Letters **115**, (2019).

[39] E. Saitoh, M. Ueda, H. Miyajima, and G. Tatara, Applied Physics Letters **88**, 18 (2006).

[40] K. Kondou, H. Sukegawa, S. Kasai, S. Mitani, Y. Niimi, and Y. Otani, Applied Physics Express **9**, 023002 (2016).

[41] Y. Huo, F. L. Zeng, C. Zhou, and Y. Z. Wu, Physical Review Applied **8**, 014022 (2017).

[42] S. Yoon, J. Liu, and R. D. McMichael, Physical Review B **93**, 144423 (2016).

[43] C. Kim, D. Kim, B. S. Chun, K.-W. Moon, and C. Hwang, Physical Review Applied **9**, 054035 (2018).

[44] C.-F. Pai, L. Liu, Y. Li, H. W. Tseng, D. C. Ralph, and R. A. Buhrman, Applied Physics Letters **101**, 122404 (2012).

[45] S. V. Aradhya, G. E. Rowlands, J. Oh, D. C. Ralph, and R. A. Buhrman, Nano Letters **16**, 5987 (2016).

[46] Y. Wang, D. Zhu, Y. Wu, Y. Yang, J. Yu, R. Ramaswamy, R. Mishra, S. Shi, M. Elyasi, K. L. Teo, Y. Wu, and H. Yang, Nature Communications **8**, 6 (2017).

[47] X. F. Zhu, M. Harder, J. Tayler, A. Wirthmann, B. Zhang, W. Lu, Y. S. Gui, and C. M. Hu, Physical Review B - Condensed Matter and Materials Physics **83**, 2 (2011).

[48] L. Pan, W. T. Soh, N. N. Phuoc, and C. K. Ong, Physica Status Solidi (RRL) - Rapid Research Letters **12**, 1800178 (2018).

[49] A. Yamaguchi, T. Ono, S. Nasu, K. Miyake, K. Mibu, and T. Shinjo, Physical Review Letters **92**, 1 (2004).

[50] M. Hayashi, L. Thomas, Y. B. Bazaliy, C. Rettner, R. Moriya, X. Jiang, and S. S. P. Parkin, Physical Review Letters **96**, 197207 (2006).

[51] M. Hayashi, L. Thomas, C. Rettner, R. Moriya, Y. B. Bazaliy, and S. S. P. Parkin, Physical Review Letters **98**, 037204 (2007).

[52] E. B. Myers, F. J. Albert, J. C. Sankey, E. Bonet, R. A. Buhrman, and D. C. Ralph, Physical Review Letters **89**, 196801 (2002).

[53] R. Lo Conte, A. Hrabec, A. P. Mihai, T. Schulz, S.-J. Noh, C. H. Marrows, T. A. Moore, and M. Kläui, Applied Physics Letters **105**, 122404 (2014).

[54] T. Moriyama, R. Cao, J. Q. Xiao, J. Lu, X. R. Wang, Q. Wen, and H. W. Zhang, Applied Physics Letters **90**, 152503 (2007).

[55] J. Holanda, O. Alves Santos, R. O. Cunha, J. B. S. Mendes, R. L. Rodríguez-Suárez, A. Azevedo, and S. M. Rezende, Physical Review B **95**, 214421 (2017).

[56] K. Ando and E. Saitoh, Journal of Applied Physics **108**, (2010).

[57] K. Ando, S. Takahashi, J. Ieda, Y. Kajiwara, H. Nakayama, T. Yoshino, K. Harii, Y. Fujikawa, M. Matsuo, S. Maekawa, and E. Saitoh, Journal of Applied Physics **109**, (2011).

[58] T. R. Mcguire and R. I. Potter, IEEE Transactions on Magnetics **11**, 1018 (1975).

[59] W. Gil, D. Görlitz, M. Horisberger, and J. Kötzler, Physical Review B - Condensed Matter and Materials Physics **72**, 1 (2005).

FIGURE CAPTIONS

FIG. 1. Schematics of (a) the Py/Pt device and (b) an equivalent electric circuit for demonstrating the magnetization switching induced by spin–orbit torque (SOT). (c) SOT-induced ferromagnetic resonance (ST-FMR) signals for $f_{MW}$ = 6, 7, 8, and 9 GHz at $\theta_H$=45°. (d) Relationship between $f_{MW}$ and $B_{res}$. The dots are experimental data, and the solid line is a fitting curve using Eq. (2). (e) Linewidth of the ST-FMR signals, $\Gamma$, obtained by fitting using Eq. (1) as a function of $f_{MW}$. The dots are experimental data, and the solid line is their linear fitting.

FIG. 2. (a) ST-FMR signals for $f_{MW}$ below 5 GHz at $\theta_H$=45°. $P_{MW}$ was 10 dBm. (b–g) Enlarged ST-FMR signals around $B_{ext}$ = 0 mT at $f_{MW}$ = (b) 10 MHz, (c) 200 MHz, (d) 1 GHz, (e) 2 GHz, (f) 3 GHz, and (g) 5 GHz.

FIG. 3. (a) ST-FMR signals for $f_{MW}$ below 5 GHz at $\theta_H$=0°. $P_{MW}$ was 10 dBm. (b–g) Enlarged ST-FMR signals around $B_{ext}$ = 0 mT at $f_{MW}$ = (b) 10 MHz, (c) 200 MHz, (d) 1 GHz, (e) 2 GHz, (f) 3 GHz, and (g) 5 GHz.

FIG. 4. $f_{MW}$ dependence of the magnitude of the hysteresis signals, $\Delta V_{FMR}$, i.e., difference between upper and lower $V_{DC}$ values of the rectangular hysteresis signals at $B_{ext}$ = 0 mT for $\theta_H$ = 0° and 45°.

FIG. 5 (a) Procedure for demonstrating SOT magnetization switching. (b) Hysteresis SOT-switching signal, $\Delta V$ = ($V_{DC2} - V_{DC1}$), as a function of pulse charge current density $J_{PLS}$ at $\theta_H$=0°. Steps (1)–(3) were carried out for every $J_{PLS}$. The raw data are shown in the inset. The main panel shows the SOT-switching signal where $V_{DC2}$ in the magnetization direction along +y was set to 0 μV for easy analysis. (c) $\Delta V$ as a function of $J_{PLS}$ at $\theta_H$=0°. Steps (1)–(3) were carried out only before starting the measurements. (d, e) $\Delta V$ as a function of $J_{PLS}$ at various values of (d) $f_{MW}$ and (e) the pulse width $W_{PLS}$ at $\theta_H=0°$. (f) $\Delta V$ and (g) normalized $\Delta V$ as functions of $J_{PLS}$ for $L_x$ = 500 nm, 2 μm, 4 μm, and 20 μm. The inset of (f) shows the magnitude of the SOT-switching signal as a function of $L_x$.

FIG. 6 (a) AMR ratio as a function of $B_{ext}$ for the Py/Pt device, obtained from two terminal resistances between the signal line and ground line in Fig. 1(b). (b) AMR ratio as a function of angle between direct current $I$ and magnetization $M$ for the Py/Pt films. The AMR ratio was calculated from the four terminal resistances.

FIG. 7 (a) $V_{DC}$–$B_{ext}$ curve for the Py electrode with $L_x$ = 100 nm and $L_y$ = 1000 nm at $\theta_H=0°$. $f_{MW}$ and $P_{MW}$ were 200 MHz and 5 dBm, respectively. (b) $\Delta V$ as a function of $J_{PLS}$ at various $B_{PLS}$ values with $W_{PLS}$ = 1 μs. (c) $V_{DC}$–$B_{ext}$ curve for the three Py electrodes in series on a Pt layer. The edge-to-edge distance between the adjacent Py electrodes was 5 μm. $L_x$, $L_y$, $f_{MW}$, and $P_{MW}$ were 500 nm, 20 μm, 200 MHz, and 10 dBm, respectively. (d) $V_{DC}$–$B_{ext}$ curve for the 4-nm-thick Co electrode with $L_x$ = 1000 nm and $L_y$ = 4000 nm at $\theta_H=0°$. $f_{MW}$ and $P_{MW}$ were 200 MHz and 5 dBm, respectively.

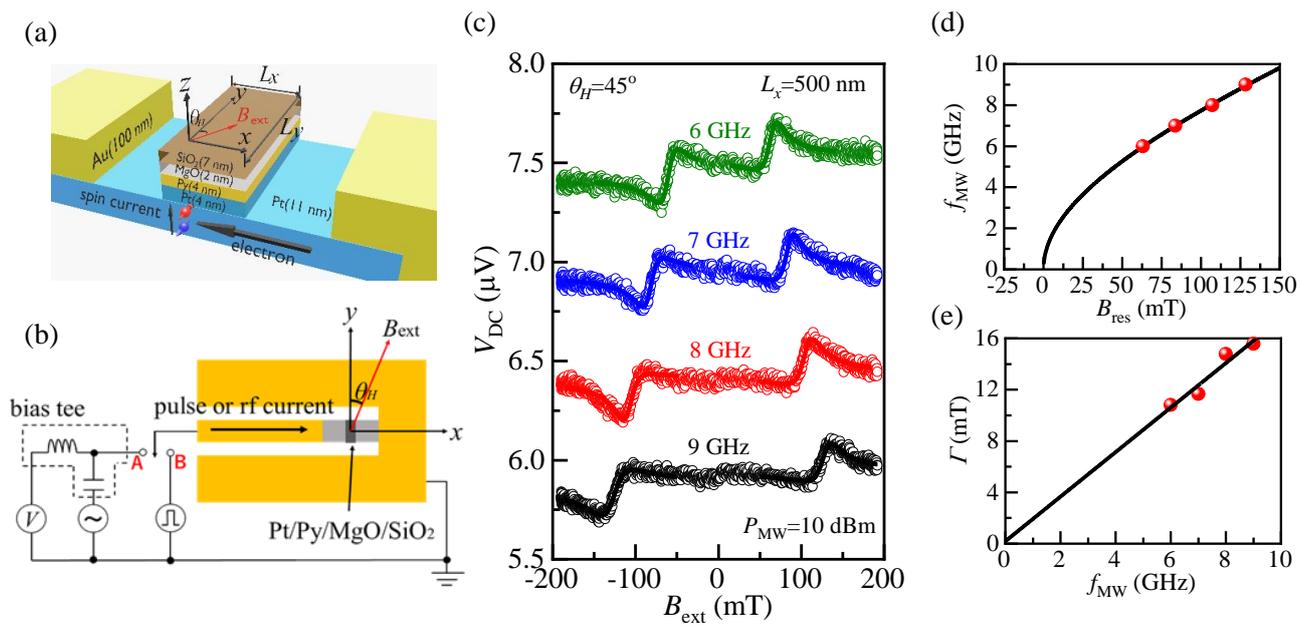

Fig. 1 M. Aoki et al.

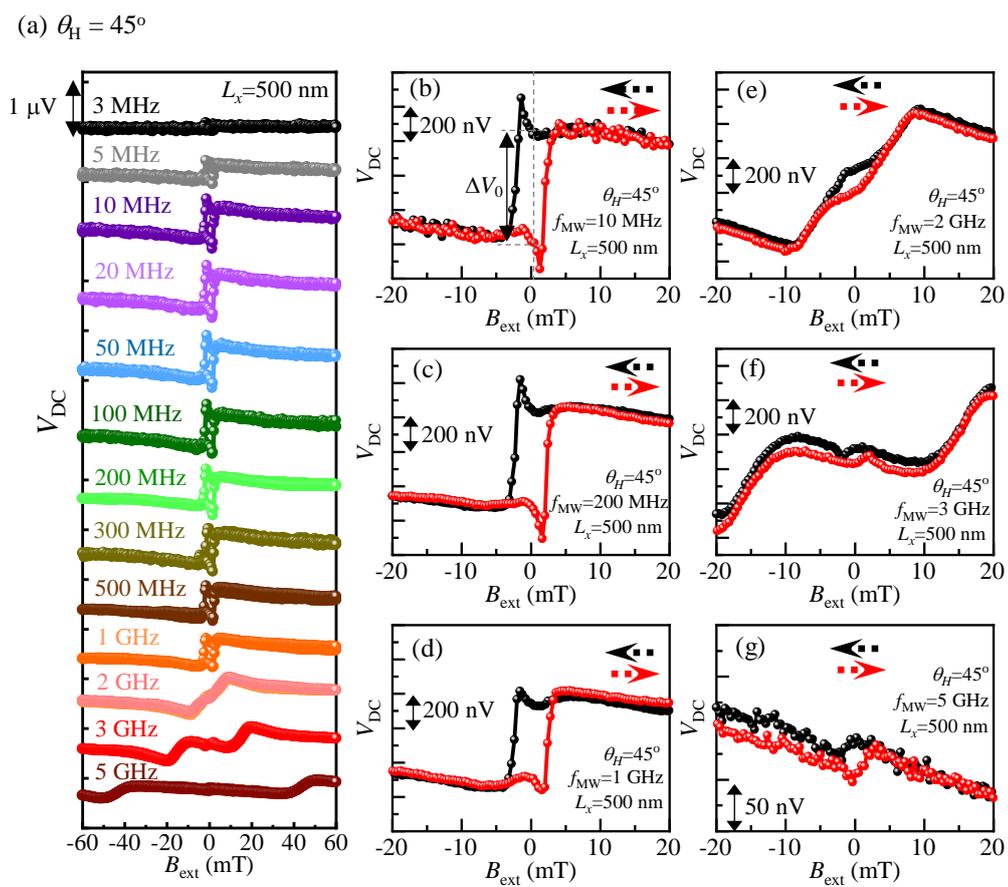

Fig. 2 M. Aoki et al.

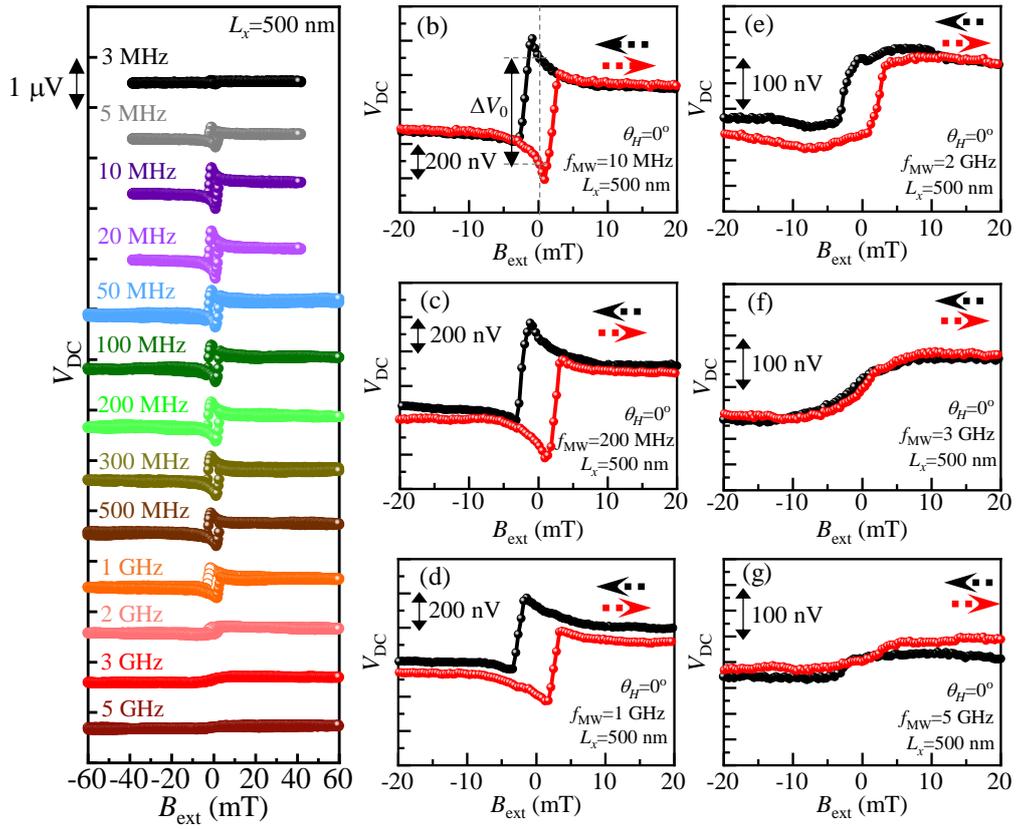

Fig. 3 M. Aoki et al.

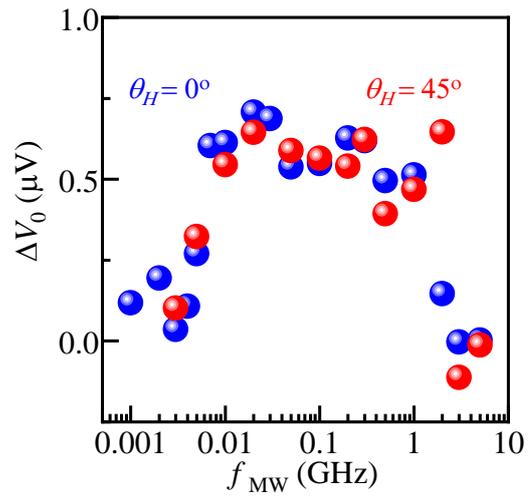

Fig. 4 M. Aoki et al.

(a)
- (1) Application of $B_{set}$ to set initial magnetization direction
- (2) Application of microwave and detection of $V_{DC1}$ as a initial value
- (3) Switch the circuit from A to B
- (4) Application of pulse current, $J_{PLS}$ at $B_{PLS}$
- (5) Switch the circuit from B to A
- (6) Application of microwave and detection of $V_{DC2}$

Repeat with different $J_{PLS}$

Optional except for the first initialization

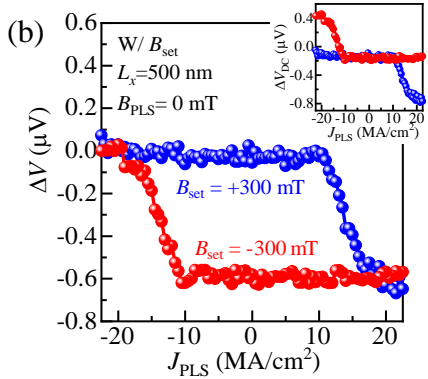
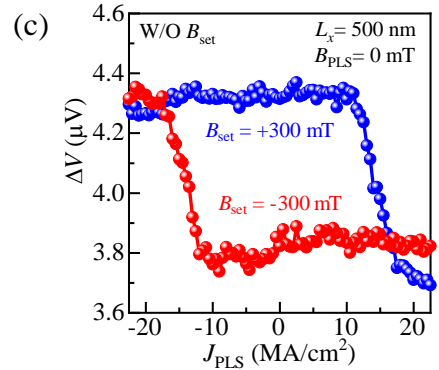
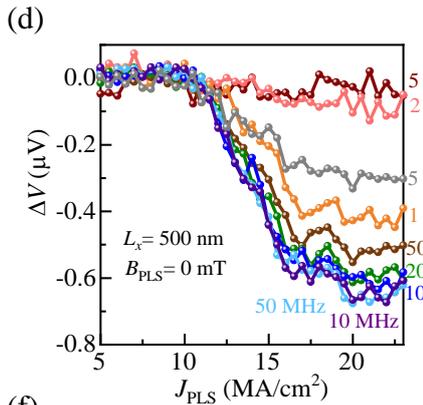
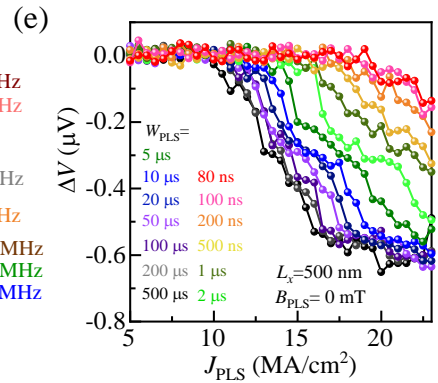
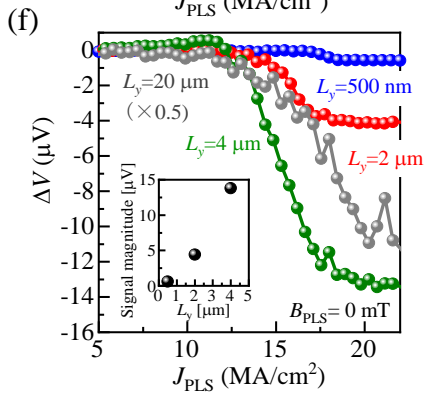
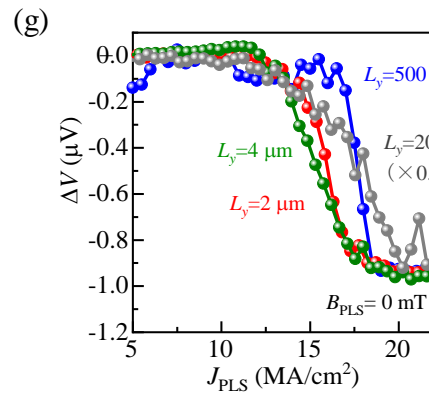

Fig. 5 M. Aoki et al.

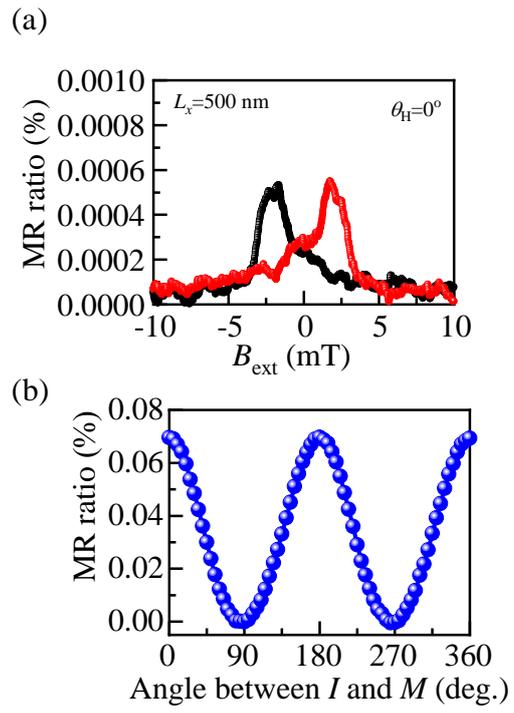

Fig. 6 M. Aoki et al.

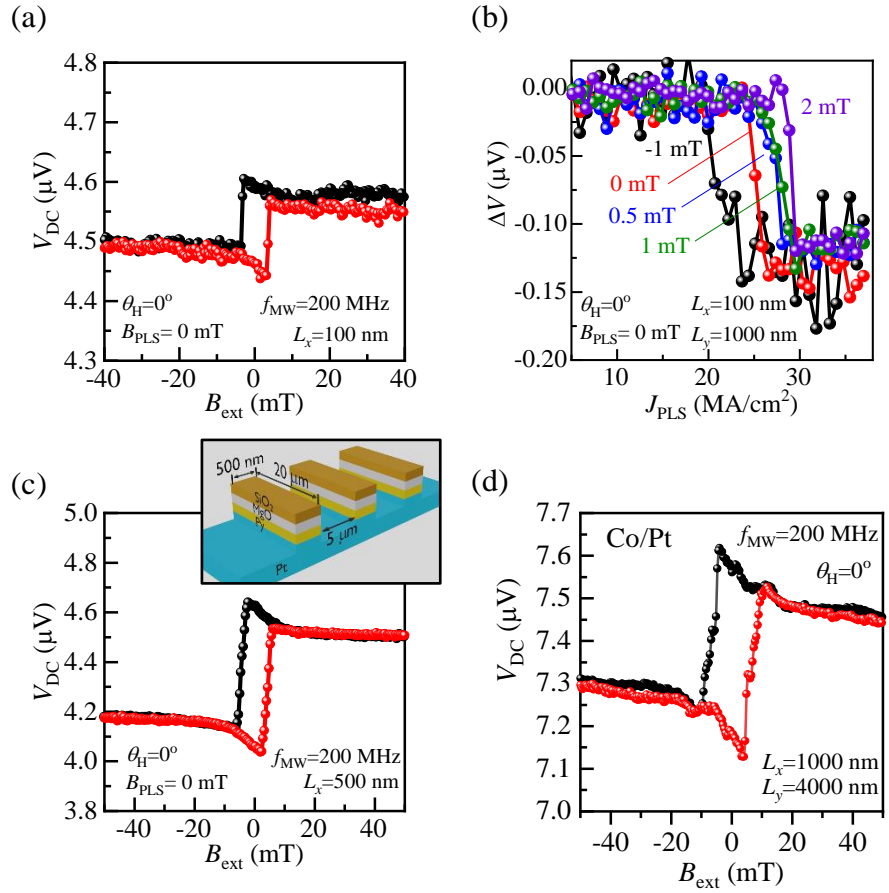

Fig. 7 M. Aoki et al.

*Supplemental Material*

# In-plane spin–orbit torque magnetization switching and its detection using the spin rectification effect at sub-GHz frequencies

Motomi Aoki[1], Ei Shigematsu[1], Ryo Ohshima[1], Honda Shyta[2], Teruya Shinjo[1], Masashi Shiraishi[1] and Yuichiro Ando[1,†]

[1]Department of Electronic Science and Engineering, Kyoto University, Kyoto, Kyoto, Japan

[2]Department of Pure and Applied Physics, Kansai University, Suita, Osaka 564-8680, Japan

**Sample fabrication procedure**

Figure S1 shows a schematic of the device fabrication procedure. First, (1) a rectangular pattern was drawn using electron beam (EB) lithography with a positive resist (PMMA/MMA double layers and ESPACER 300Z). Then, (2) MgO (2 nm) / $Ni_{80}Fe_{20}$ (Py, 4 nm) / Pt (15 nm) layers were deposited using ultra-high-vacuum EB deposition. The base pressure was $5\times10^{-7}$ Pa, the deposition pressure was ~$5\times10^{-6}$ Pa, and the deposition rates were 0.15 Å/s for Pt, 0.10 Å/s for Py, and 0.05 Å/s for MgO. The sample was exposed to the air, and then a 7-nm-thick $SiO_2$ layer was deposited on the MgO layer via RF magnetron sputtering with a base pressure of $3\times10^{-5}$ Pa, deposition pressure of 0.5 Pa, deposition power of 100 W, and deposition rate of 0.37 Å/s. After liftoff of the $SiO_2$/MgO/Py/Pt layers, (3) a rectangular ferromagnetic electrode was drawn using EB lithography with a negative resist (NEB-22 and ESPACER AX01). Then, (4) argon ion ($Ar^+$) milling was carried out and stopped after milling of the 4-nm-thick Pt layer to keep an 11-nm-thick Pt channel. Finally, (5) a Au (100 nm) / Ti (3 nm) coplanar wave guide was fabricated using EB lithography (PMMA/MMA double layers and ESPACER 300Z) and EB deposition.

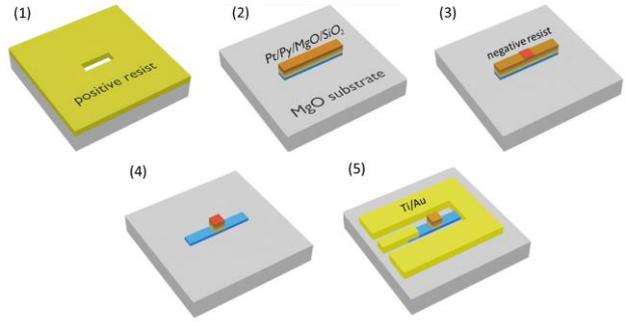

Fig. S1 A schametic of the device fabrication procedures.

**B| Micromagnetic simulation under microwave irradiation using Mumax$^3$ calculation**

The micromagnetic simulation was performed using the MuMax$^3$ package, which provides GPU-accelerated calculation of magnetization dynamics. The motion of the magnetic moment was simulated in a 500 nm × 5 μm × 4 nm rectangular geometry. The size of the unit cell was 16 nm × 39 nm × 2 nm. To excite the motion of the magnetic moment, the static magnetic field was first applied at $\theta_H = 6°$, where $\theta_H$ is the angle to the +y direction. A tilted magnetic field was used to make the domain-wall structure most stable, which enables the system to reach the steady state quickly. We confirmed that even at $\theta_H = 6°$ the steady state under microwave irradiation was the same as that at $\theta_H = 0°$. After reaching the energy-minimum state via the relax() command, an alternating magnetic field was applied. The direction of each spin in the unit cell was outputted to the files every $0.04T$ s, where $T$ is the period of the AC magnetic field.

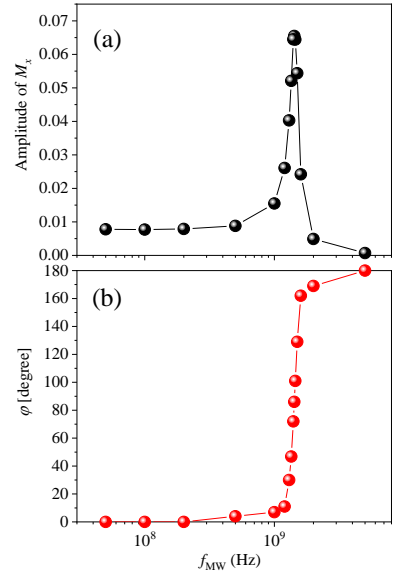

Fig. S2 Oscillation amplitude of *x*-component of magnetization $M_x(t)$ and phase difference between AC magnetic field, $\varphi$, as a function of the microwave frequency, $f_{MW}$.

We carried out the micromagnetic simulation using the typical parameters of permalloy ($Ni_{80}Fe_{20}$). The saturation magnetization $M_s$ was set to $4.6 \times 10^5$ A/m, which was obtained from the frequency dependence of the resonance field $B_{res}$ of the ST-FMR spectra (see Fig. 1(d) in the main text). The Gilbert damping constant and the exchange stiffness were set to $0.01$[1] and $1.3 \times 10^{-11}$ J/m[2], respectively. Figure S2 shows the oscillation amplitude of $M_x(t)$ and its phase difference with the AC magnetic field, $\varphi$, as a function of frequency $f_{MW}$. The quantity $M_x(t)$ is the $x$ component of the net magnetization of all unit cells. $M_x(t)$ was maximized and $\varphi$ became $90°$ around 1.425 GHz, indicating resonance. In contrast with $M_x(t)$ for $f_{MW} > 2$ GHz, a considerable $M_x(t)$ is still obtained for $f_{MW} < 1$ GHz.

## C| Micromagnetic simulation including SOT

Section B revealed that $\varphi$ is $0°$ for $f_{MW} < 1$ GHz. In this situation, the SOT contribution to $\Delta V_0$ is expected to be negligible. However, because the external magnetic field was 0 mT during the LFST-FMR measurements, the magnetization direction at each position has a considerable dispersion especially near the edge area, resulting in a dispersion of the FMR condition. This means some positions match the FMR condition even for $f_{MW} < 1.425$ GHz, and enhancement of the precession angle is expected. In this case, $\varphi$ also changes, and the SOT contribution to $\Delta V_0$ becomes pronounced. Therefore, we carried out micromagnetic simulations that considered the SOT contribution. The time dependence of the SOT-induced magnetic moment in the Py film was calculated using the micromagnetic simulation [4]. The Py wires used each had a length of 20 μm in the $y$ direction, width of 500 nm in the $x$ direction, and thickness of 5 nm in the $z$ direction. The wire was divided into small cubic cells with dimensions of 5 nm × 5 nm × 5 nm. The effective magnetic field was composed of long-range magnetic dipole–dipole interactions and short-range exchange interactions among the neighboring computational cells. The parameter values were chosen according to the properties of the Py: the damping constant was set to 0.05, $M_s$ was 460 kA/m (= 0.58 T), and the exchange stiffness constant $A$ between the adjacent magnetic moments was 13.0 pJ/m. The spin injection velocity and spin diffusion length for the SOT were set to $u = 5$ m/s and 3 nm, respectively. The spin current was injected into the Py electrode via the bottom surface.

Figure S3 shows $M_y(t)$ as a function of $f_{MW}$. We calculated $M_y(t)$ by averaging the magnetization of the whole ferromagnetic area. A considerable $M_y(t)$ appeared mainly near the edge area because of the large spin torque due to the tilted magnetization. $M_y(t)$ was maximized around $f_{MW} = 1.5$ GHz, consistent with the results in Section B. Note that a considerable $M_y(t)$ was obtained even for $f_{MW} < 1$ GHz and obviously larger than $f_{MW} = 0$ GHz.

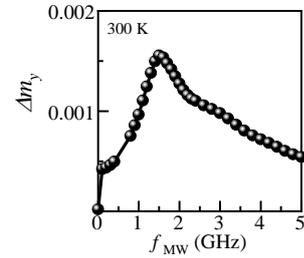

Fig. S3 Oscillation amplitude of $y$-component of magnetization $M_y(t)$ indeced by SOT as a function of the microwave frequency, $f_{MW}$.

## D| Calculation of DC voltage via the spin rectification effect

Using the above results, we calculate the DC voltage under irradiation with low-frequency microwaves. $V_{DC}$ can be described by

$$V_{DC} = <I_{MW}\sin(2\pi f_{MW}t) \cdot R(t)>_t, \quad (1)$$

where $I_{MW}$ and $R(t)$ are the rf current and the resistance of the Pt/Py bilayer, respectively. $<>_t$ means the time average. The resistance $R(t)$ including AMR is described as $R = R_0 + \Delta R\sin^2\theta_M(t)$, and $M_x(t) = \sin\theta_M(t)$ because $M_x(t)$ is normalized. Therefore,

$$V_{dc} = \frac{5\times 10^{-6}}{L_y} I_{rf}\Delta R < \sin(2\pi f_{MW}t) \cdot M_x^2(t)>_t, \quad (2)$$

where $\theta_M(t)$ is the angle between the magnetization $M$ and the $y$ axis, $R_0$ is the resistance at $\theta_M = 0°$, $\Delta R$ is the difference in resistance between $\theta_M = 0°$ and $\theta_M = 90°$, and $L_y$ is the length of Py along the $y$ axis in the measurement. The expression is multiplied by $\frac{5\times 10^{-6}}{L_y}$ because most of the contribution to the motion of $\boldsymbol{M}(t)$ is the edge area of the rectangle as shown in Fig. S4, where $\boldsymbol{M}(t)$ is the magnetization vector. We note that $M_x^2(t)$ is not the square of $M_x(t)$ but the spatial average of the square of the magnetic moment in each unit cell. We adopted $M_x^2(t)$ at 200 MHz. In this case, $<\sin(2\pi ft) \cdot M_x^2(t)>$ was $1.5\times 10^{-3}$. $I_{MW}$ and $\Delta R$ were $1.8\times 10^{-2}$ A and $4.9\times 10^{-2}$ Ω, respectively, in the measurement of the 500 nm × 20 μm sample, so $V_{DC}$ was calculated to be $3.4\times 10^{-9}$ V. The calculated $V_{DC}$ was two orders of magnitude lower than the measured one. One possible origin is the difference in the temperature. The temperature was set to 0 K in the simulation, which restricted the motion of magnetization.

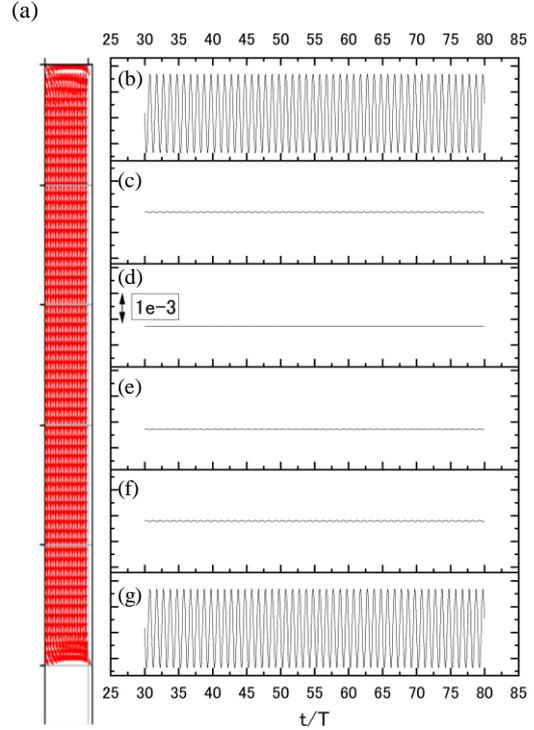

Fig. S4 Magnetic domain structure of Py electrode with $L_x$=500 nm and $L_y$=20 μm at zero magnetic field. Oscillation of $m_y$ of each part of the Py electrode under application of a microwave with $f_{MW}$=200 MHz. The Py electrode was divided equally into six parts and net of mgnetization of each part was shown. The distance of the each part from the edge of the electrodes was (b, g) 0~0.83 μm, (c, f) 0.83~1.67μm (e, d) 1.67~2.5μm, respectively.

## E| Calculation of $V_{ISHE}$

The spin current $j_s$ can be calculated from spin pumping via

$$j_s = \left(\frac{5}{L_y}\right)^2 f_{MW} \int_0^{1/f} \frac{\hbar}{4\pi} g_r^{\uparrow\downarrow} \left[\bm{M}(t) \times \frac{d\bm{M}(t)}{dt}\right]_y dt, \quad (3)$$

and the DC voltage $V_{ISHE}$ can be calculated from the inverse spin Hall effect via

$$V_{ISHE} = L_y R \theta_{SHE} \lambda_{Pt} \tanh\left(\frac{d_{Pt}}{2\lambda_{Pt}}\right)\left(\frac{2e}{\hbar}\right) j_s, \quad (4)$$

where $\hbar$, $d_{Pt}$, and $e$ are the Dirac constant, thickness of the Pt layer, and elementary charge, respectively. We used $\bm{M}(t)$ when $f = 200$ MHz. The $x$ and $z$ components of $\bm{M}(t)$ were $5.2 \times 10^{-2} + 4.9 \times 10^{-2} \sin 2f\pi t$ and $6.0 \times 10^{-5} \cos 2\pi ft$, respectively. Assuming the typical mixing conductance $g_r^{\uparrow\downarrow} = 2.31 \times 10^{19}$ $m^{-2}$, spin Hall angle $\theta_{SHE} = 0.04$, and spin diffusion length $\lambda_{Pt} = 10$ nm, $j_s$ and $V_{ISHE}$ were calculated to be $1.7 \times 10^{-14}$ J/m² and $2.4 \times 10^{-13}$ V, respectively, for $L_y = 20$ μm and $d_{Pt} = 15$ nm. This value of $V_{ISHE}$ is negligible compared with $V_{DC}$ calculated in section C. Even if we assume the resonance condition ($f = 1.425$ MHz), $V_{ISHE}$ is calculated to be $6.4 \times 10^{-10}$ V, which is also negligible.

## References

[1] G. D. Fuchs et al., Appl. Phys. Lett. **91**, 062507 (2007).

[2] A. Vansteenkiste et al., AIP Adv. **4**, 107133 (2014).

[3] K. Ando et al., J. Appl. Phys. **109**, 103913 (2011).

[4] S. Honda and H. Itoh, J. Nanosci. Nanotech. **12**, 8662 (2012).